\documentclass{mem}
\usepackage{natbib}\usepackage{txfonts}\usepackage{balance}
\usepackage{graphicx}
\usepackage[a4paper]{hyperref}
\idline{75}{282}
\begin{document}
\def\teff{$T\rm_{eff }$}
\def\kms{$\mathrm {km s}^{-1}$}

\title{
X-ray variability with WFXT
}

   \subtitle{AGNs, transients and more}

\author{
M. \,Paolillo\inst{1,8} 
\and C.\, Pinto\inst{2,1}
\and V.\, Allevato\inst{3,1}
\and D.\, de Martino\inst{4}
\and M.\, della Valle\inst{4}
\and I.\, Papadakis\inst{5}
\and R.\, Gilli\inst{6}
\and P.\, Tozzi\inst{7}
\and the WFXT collaboration
          }

  \offprints{M. Paolillo}

\institute{
Universit\`a Federico II, Dip. di Scienze Fisiche
C.U. Monte S.Angelo, Napoli, Italy
\email{paolillo@na.infn.it}
\and
SRON, Sorbonnelaan 2, 3584 CA Utrecht, the Netherlands
\and
Max-Planck-Institut f\"ur Plasmaphysik, Boltzmannstrasse 2, D-85748 Garching, Germany
\and
Istituto Nazionale di Astrofisica --
OAC, Napoli, Italy
\and
University of Crete Dept Physics, P.O. Box 2208, GR 710 03 Heraklion, Greece
\and
Istituto Nazionale di Astrofisica --
OABo, Bologna, Italy
\and
Istituto Nazionale di Astrofisica --
OATs, Trieste, Italy
\and
Istituto Nazionale di Fisica Nucleare, Napoli, Italy
}

\authorrunning{Paolillo et al.}

\titlerunning{X-ray variability with WFXT}

\abstract{
The Wide Field X-ray Telescope (WFXT) is a proposed mission with a high survey speed, due
to the combination of large field of view (FOV) and effective area, i.e. grasp, and sharp
PSF across the whole FOV. These characteristics make it suitable to detect a large number
of variable and transient X-ray sources during its operating lifetime. Here we present estimates
of the WFXT capabilities in the time domain, allowing to study the variability of thousands of AGNs with
significant detail, as well as to constrain the rates and properties of hundreds of distant, faint and/or rare 
objects such as XRF/faint GRBs, Tidal Disruption Events, ULXs, Type-I bursts etc.
The planned WFXT extragalactic surveys will thus allow to trace variable and transient X-ray populations 
over large cosmological volumes.
}
\maketitle{}

\section{Introduction}
The ability to conduct timing studies has always characterized X-ray astronomy, but  
so far, due to the limited sensitivity and field-of-view (FOV) of the instruments on board of X-ray satellited, 
studies were concentrated on individual and relatively nearby and bright sources.

The Wide Field X-ray Telescope (WFXT) is a proposed X-ray mission characterized by a wide field (1 square degree), a large effective area ($1 m^2$ @ 1 keV) and a costant PSF across the entire FOV (goal design, see Rosati et al. in this volume). While not designed to be a monitoring mission, its capabilities and the proposed observing strategy, make it suitable to conduct timing studies for an unprecedented number of moderate and high redshift AGNs, as well as to discover and constrain rates and properties of distant, faint and rare X-ray populations such as X-ray Flashes/faint GRBs, Tidal Disruption Events, ULXs, Type-I bursts etc.

In this work we present estimates of the WFXT monitoring capabilities for AGNs and other variable/transient sources, that can be detected in the 3 main extragalactic surveys planned for the mission. In this respect the work presented here represents the minimum achievements expected by WFXT in the time domain, since more specific and/or dedicated studies (e.g. galactic surveys, nearby galaxies monitoring etc.) will certainly increase the WFXT impact in the field.

\section{Monitoring SMBH accretion with WFXT}
Monitoring campaigns on nearby galaxies have shown that intense variability on all timescales, from hours to years, is a common property of all AGNs. This variability increases with energy,
and is very intense in the X-ray regime, in close resemblance with the one observed in galactic accreting Black Holes (BH)  \cite[see][for a comprehensive review]{McHardy10}. 
Long observations, as those required to conduct deep surveys (\textit{Chandra} Deep Fields, Lockmann Hole), allowed to study variability also in higher redshift sources, confirming that variability is common to all AGNs over cosmological volumes, and that it reflects the details of the accretion process and the properties of the system (mass, accretion rate, obscuration) \citep{Almaini00,Paolillo04}.
X-ray variability can thus be used as a tool to trace the accretion history of SMBH across cosmic time \citep{Papadakis08,Allevato09}. Such attempts however have been hampered by the random sampling pattern and small number of sources accessible with the present generation of X-ray satellites. 

The study of AGN populations is one of the primary objectives of the WFXT missions. In its first 5 years WFXT will conduct 3 extragalactic surveys that are predicted to detect an unprecedented number of AGNs ($>10^7$, see Gilli et al. and Matt \& Bianchi contributions in this volume). While a large number of these sources will be close to the detection limit, the WFXT grasp (Rosati et al., this volume) ensures that a considerable fraction will be detected with several thousand photons, thus allowing to perform variability studies on the temporal baselines sampled by the different surveys.

\subsection{Predicting the number of variable AGNs}
\label{simulations}
\begin{figure*}[]
 {\centering
 \resizebox{0.8\hsize}{!}{\includegraphics{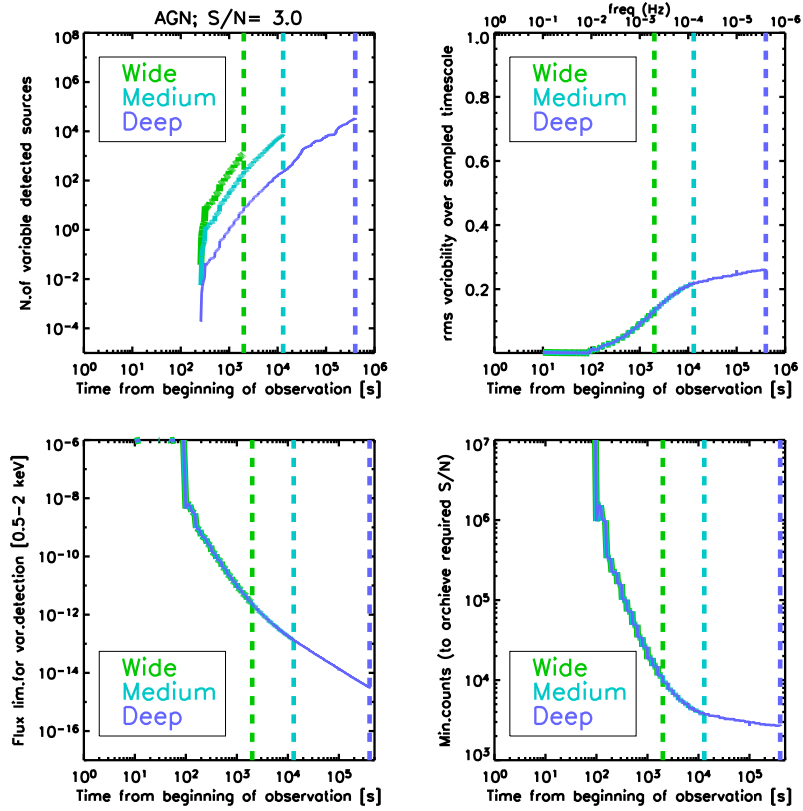}}
 \caption{
 \footnotesize AGN variability estimates for the WFXT extragalactic surveys as a function of the sampled timescale: \textit{Top panels}) Number of sources for which variability is detected with the required S/N (left) and average rms variability detected at every timescale (right); \textit{Bottom panels}) Flux limit for variability detection within the assumed S/N threshold (left) and corresponding total counts (right). The vertical dashed lines represent the duration limit of each survey, assuming continuous monitoring.
 \label{AGN_var}}}
 \end{figure*}

In order to evaluate the WFXT monitoring capabilities for AGNs, we initially assumed that each field in the 3 planned surveys will be observed continuously. While this is feasible for the {\it Wide} and {\it Medium} survey due to its short exposure time per field (2 ks and 13 ks), it is unrealistic for the {\it Deep} survey due to the long exposure times (400 ks) and the visibility constraints for a low-orbit mission; we will discuss how to relax these constraints in \ref{strategy}. 
To simulate the intrinsic AGN variability we adopted a template X-ray Power Density Spectrum (PDS) observed in nearby AGNs (e.g. NGC 4051) displaying the characteristic power-law shape with a high-frequency break \cite[e.g.][]{Uttley02}. We further required  at least 10 bins of equal duration in the X-ray lightcurve, and that the average signal-to-noise ratio (S/N) per bin in the satellite band, due to the intrinsic variability, is larger than a fixed threshold ($S/N=3$ in Figure \ref{AGN_var}). Note that these requirements are much more stringent than a simple variability detection, allowing to derive constraints on the underlying physical processes.  
We adopted the background estimates of Tozzi et al. (this volume) which include particle contribution, galactic ISM and unresolved AGNs/galaxy clusters. 
The expected number of variable AGNs was finally derived from AGN number counts \citep{has95,Giacconi02,Luo08}, after converting the S/N limit into a flux limit, and multiplying for the total angular coverage of each survey (20000, 3000 and 100 sq.deg. for the Wide, Medium and Deep survey respectively). 

The results of the simulations are shown in Figure \ref{AGN_var}. As the sampled timescale increases, the flux and count limits for a sound variability detection decrease, due to a combination of longer integration time and a larger intrinsic rms variability produced by the power-law behavior of the AGN PDS. The three surveys will reach limiting (variability) sensitivities, for $S/N>3$, of $\sim 2\times 10^{-12},~1\times 10^{-13 }~\mbox{and}~5\times 10^{-15}$ erg/s for the Wide, Medium and Deep survey respectively in the 0.5-2 keV band. With several thousand of counts in each source this sample largely overlaps with the one suited for spectroscopic studies (Gilli et al., this volume), allowing a detailed characterization of these AGNs. The predicted rms range between 10\% and 25\%, in good agreement with the values reported for deep extragalactic surveys \cite[see e.g.][]{Paolillo04}. While the Wide survey covers a much larger sky area, the longer integration will favour the Medium and Deep surveys, both in terms of accessible flux range and variability timescales.
 
 \begin{figure}[]
 \resizebox{\hsize}{!}{\includegraphics{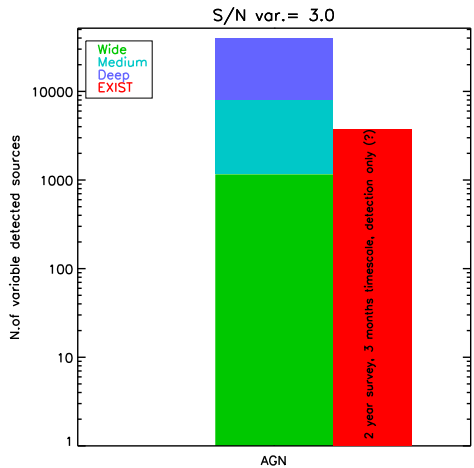}}
 \caption{
 \footnotesize Number of AGNs for which WFXT is expected to detect variability with $S/N>3$, in each of the 3 planned extragalactic surveys. For comparison we show an estimate of the number of AGNs that will be detected by the EXIST mission, during the 2 years survey phase, with $>10$ data points, as extrapolated from \cite{DellaCeca09}. 
 \label{AGN_histo}}
 
 \end{figure}

These results are summarized in Figure \ref{AGN_histo}, where we present the WFXT performance in terms of the number of AGNs with variability detected at the $>3\sigma$ level, over the whole lightcurve. 
In this respect the WFXT capabilities appear comparable to other planned missions with a more stringent monitoring character: for instance the wide field EXIST mission \citep{Grindlay07} will also monitor thousands of AGNs but, due to the different energy band, spatial resolution, sensitivity and sampling pattern, will offer a complementary view of more nearby AGNs \citep{DellaCeca09}.

\subsection{Constraining the observing strategy}
\label{strategy}
\begin{figure*}[]
 \resizebox{\hsize}{!}{\includegraphics{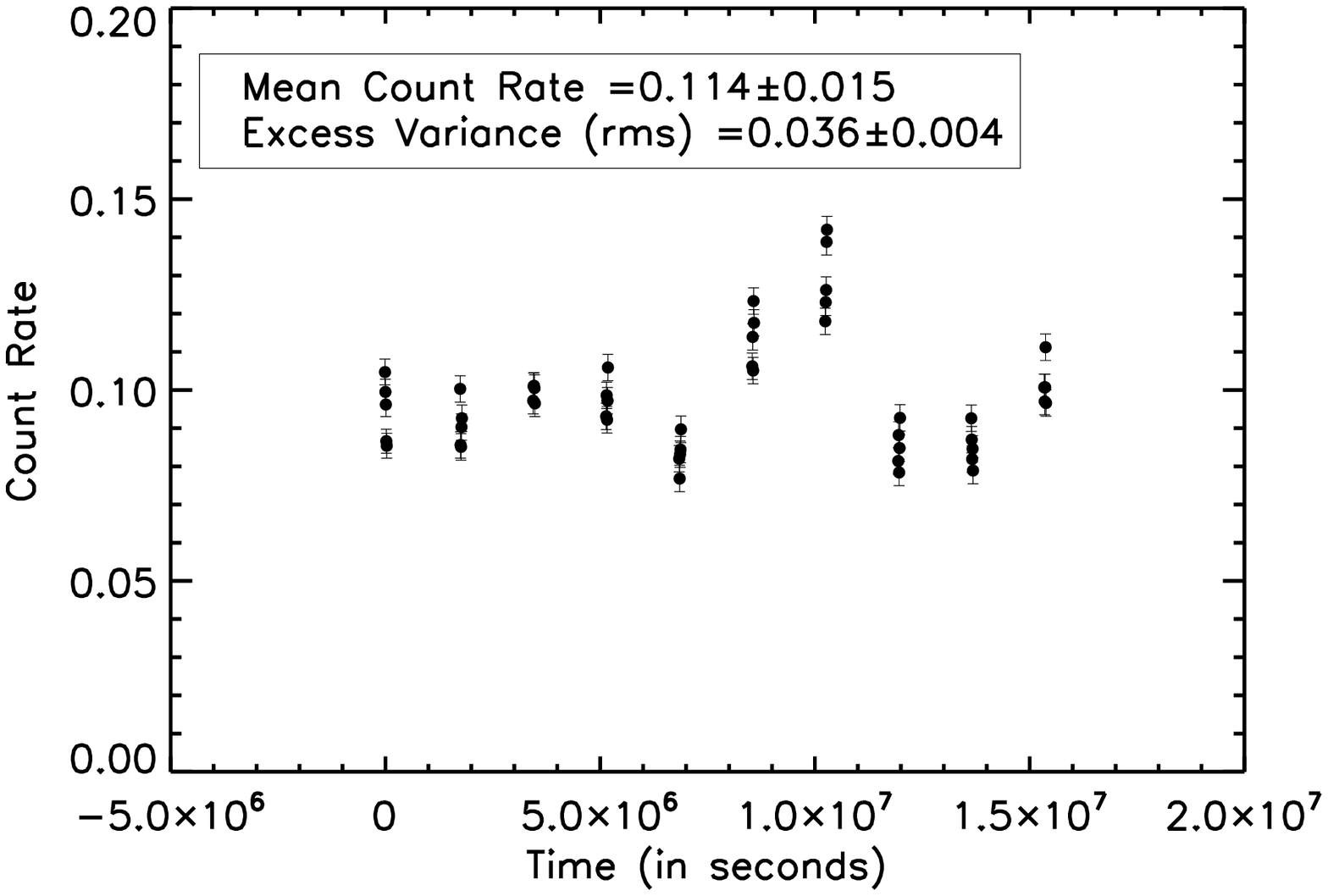}\includegraphics{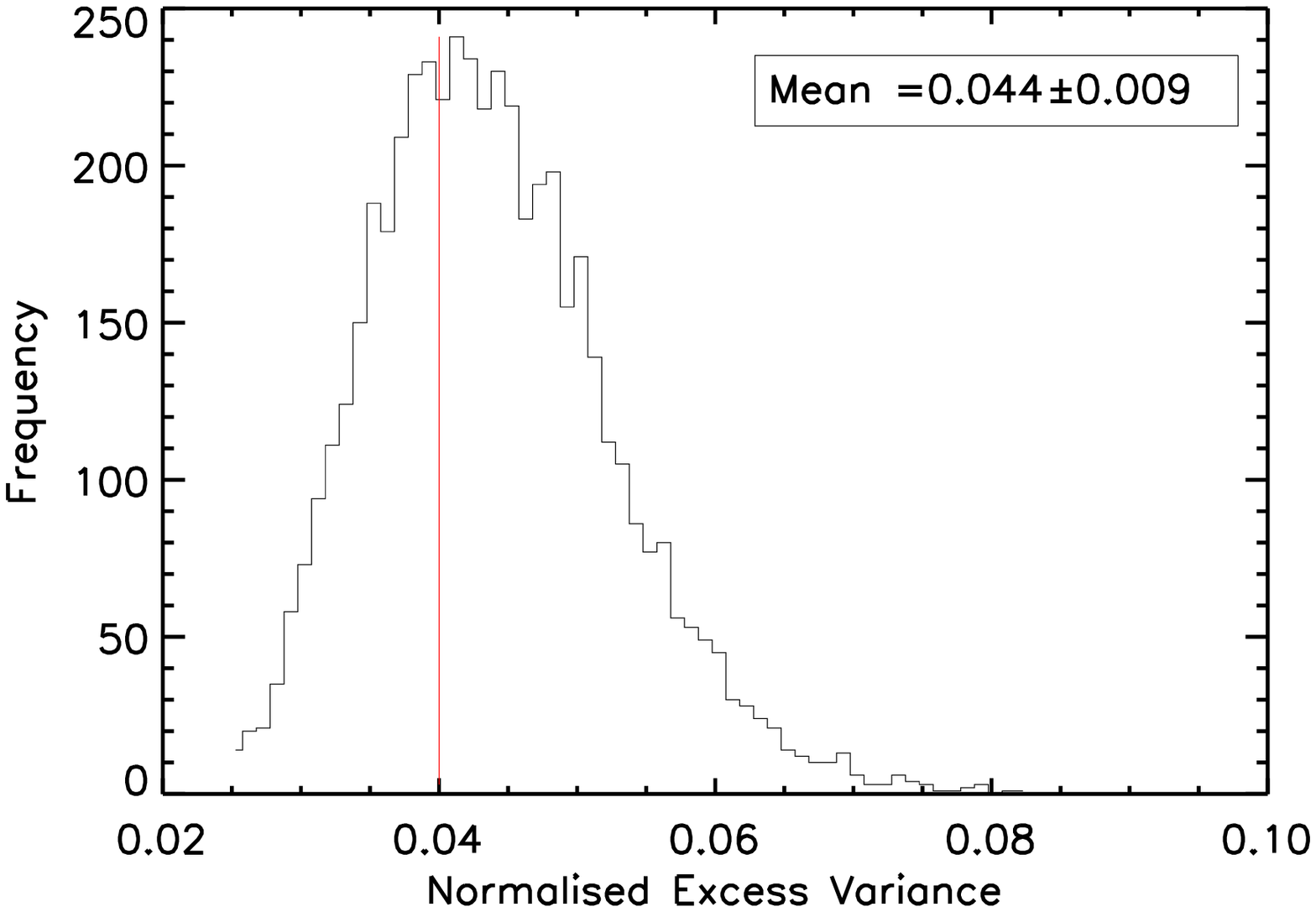}}
 \caption{
 \footnotesize \textit{Left:} One example of a simulated WFXT AGN lightcurve for the Deep survey, reproducing flux and variability of one of the CDFS sources \citep{Allevato09}, sampled in 50 ks observations spread uniformly over $\sim 6$ months. \textit{Right:} The \textit{excess variance} distribution of 5000 simulated lightcurves, such as the one shown in the left panel, compared to the input value (vertical line) fixed at 0.04, i.e. 20\% rms. This observing strategy will be able to retrieve the intrinsic variance with a $1\sigma$ uncertainty of $21\%$. 
 }
 \label{sampling}
 \end{figure*}

The study of X-ray variability in nearby AGNs has shown that the AGN PDS depends most likely on the mass of and accretion rate onto the central supermassive BH \citep{mchardy06}. The expectations based on the template PDS spectrum used in \S \ref{simulations} may not be realistic when we assume a wide range of masses and accretion rates: for more massive/lower accreting sources the expected variability will be lower than assumed so far. 
On the other hand the continuous monitoring hypothesis must also be relaxed, thus compensating the reduced variability with longer temporal baselines. To investigate the impact that a sparse sampling pattern may have on the final variability measurements, we performed a set of Monte Carlo simulations of AGNs lightcurves in order to quantify the bias of the variability estimator. To this end we modified the original \cite{Timmer95} algorithm that generates red-noise data with a power-law density spectrum, to reproduce the sampling pattern, background and sensitivities expected by WFXT. Figure \ref{sampling} (left panel) presents an example of a possible observing scheme for a single field in the Deep survey, where observations of 50 ks each are spread evenly over $\sim 6$ months. Such scheme would allow to derive AGNs lightcurves with gaps of a few weeks, thus sampling both high and low frequencies in the PDS. Additionally such pattern could be useful to discover and trace transients with long decay timescales such as Tidal Disruption Events which are believed to be due to stellar disruptions near quiescent SMBH (see \S \ref{transients_sec}). In Figure \ref{sampling} (right panel) we show the excess variance (i.e. the fractional variability) distribution for 5000 simulations of the sparsely sampled lightcurve compared to the input value. The retrieved mean excess variance agrees well with the input value, while the uncertainty on the single measurements is of the order of 20\%. The large AGN samples provided by WFXT will allow to further reduce the statistical uncertainties on each of the studied subsamples.
Finally this observing pattern will allow to extend the sampled frequency range, thus making the observation suitable to study the bulk of the AGN population down to low accretion/higher mass SMBHs.

\section{Transient and variable sources}
\label{transients_sec}

Other than AGNs, a large variety of variable and transient sources can be predicted to be observed by WFXT during its operating lifetime. To provide some quantitative estimates of the WFXT capabilities in this field, we took into consideration a few of the most likely variable objects that will be observed during the WFXT main surveys: Tidal Disruption Events (TDEs), Low Luminosity GRBs/X-ray flashes (LL-GRB, XRF), Super Soft Sources (SSS), Ultraluminous X-ray binaries (ULX), Type I bursts etc.

\begin{figure*}[t]
 \centering{\resizebox{0.7\hsize}{!}{\includegraphics{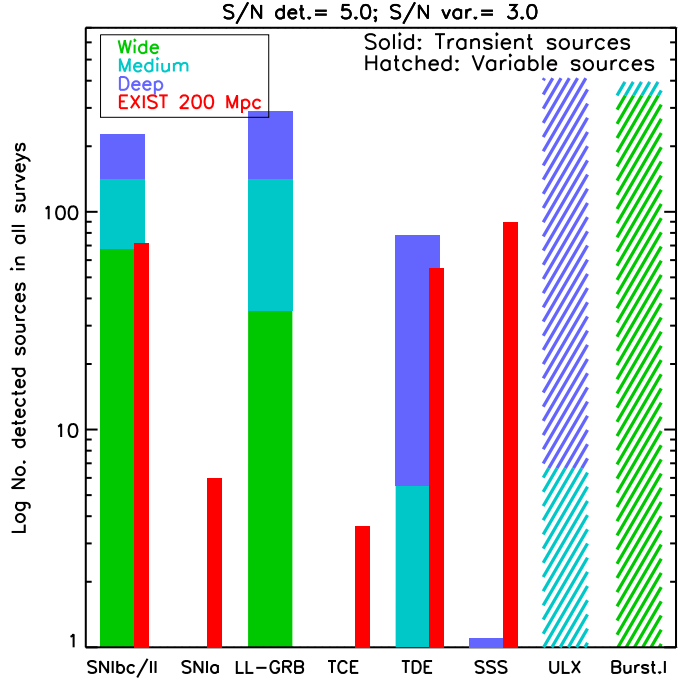}}}
 \caption{
 \footnotesize Number of transient/variable sources expected in the WFXT extragalactic surveys. Recurring bursters are reported as hatched bars. For comparison we show the expected numbers within 200 Mpc from the EXIST mission over a comparable mission lifetime \citep{Soderberg09}.
 }
 \label{transients_histo}
 \end{figure*}

 \noindent\textbf{TDE:} are believed to occur when a star is disrupted in the proximity of a quiescent SMBH, fueling the BH and revealing itself through the UV and X-ray emission due to the gravitational energy release of the accreted material. TDE are one of the few means to detect quiescent SMBH in distant galaxies. So far however only a handful of events have been serendipitously observed \citep[e.g.][]{Gezari09}. The large area covered by WFXT will make the detection of such events very likely, especially since their emission is expected to peak in the FUV/soft X-ray bands.\\
\noindent\textbf{LL-GRBs/XRF:} X-ray bursts have been detected in coincidence with SN explosions and/or GRBs. The fenomenology of such events is still poorly understood and both precursor type and rates are very debated in the literature, due to the small number of serendipitous detections observed so far. In this work we took into consideration two possible types of X-ray transients associated with SN events: XRF and LL-GRBs. The first class has its template in XRF 080109 observed by the Swift satellite without an associated GRB, and its origin can be due to the breakout of either the SN shock \citep{Soderberg08} or a mildly relativistic jet \citep{Mazzali08}. On the other end LL-GRB, such as GRB 060218 \citep{Campana06}, could represent the X-ray counterpart of many associated GRB-SNe. The number of future detections (see Table \ref{transient_tab}) depends on both the intrinsic SN rates \citep[e.g.][]{Cappellaro99} and the opening angle of the associated jet \citep{Guetta07}.\\
\noindent\textbf{SSS:} Supersoft sources are X-ray emitters detected at energies below 1 keV, with X-ray luminosities of $\rm 10^{36\div 38}\,erg\,s^{-1}$, and characterized by blackbody-like spectra with temperatures of $15-80$ keV \citep{Kahabka06}. Believed to be mostly hydrogen-burning white dwarfs, they are found both in early and late-type galaxies. SSS have complex time variability which is irregular over hours to years.\\
\noindent\textbf{ULX:} Ultraluminous X-ray binaries are varible accreting systems whth luminosities $>10^{39}$ erg/s, i.e. larger than the Eddington luminosity for a neutron star or $5 M_\odot$ BH, which display both long-term (days-months) variability and X-ray flares on timescales of hours. They tend to be associated to star forming regions, and proposed as candidates for intermediate ($>100 M_\odot$) mass BHs \citep{Fabbiano06}. The interest in such objects is due to the difficulties in explaining their formation in standard star-formation scenarios. \\
\noindent\textbf{Type I bursts:} these are accreting neutron stars (NS) in low-mass X-ray binaries 
(LMXB) displaying  rapid (tens to hundreds of seconds) bursts with X-ray intensity many times brighter than the 
persistent level. The burst X-ray spectrum is generally consistent with a blackbody with color temperature 
of 2-3 keV, reaching X-ray luminosities up to $\rm 10^{38}\,erg\,s^{-1}$. These events are 
caused by unstable nuclear burning on the surface of the NS \citep{Galloway08}. The X-ray bursts are often recurrent on timescales from 30\,min to a few hours.\\
\noindent\textbf{Other:} many more types of rare and/or faint variable and transient sources are not discussed here in detail. For instance a breakout X-ray flash is predicted in SNIa events, while flares due to tidal compression (TCE) in stars accreted by a SMBH could mark the onset of the disruption process, triggering prompt followups. We included these in our simulations in order to cover the plausible parameter space, and highlight the WFXT capabilities as a function of the properties of the transient event.

\begin{table*}
%\begin{center}
\caption{\footnotesize Number of transients expected in the WFXT extragalactic surveys, along with the physical parameters used in the simulations and the calculated distance limits.\label{transient_tab}}
 \begin{tabular}{lcccccl}
 \hline\hline
 Type & N.sources & $L_X^{burst}$ & $L_X^{quiescent}$ & characteristic & Rate & Dist.limit \\
     & &  ($10^{40}$ erg/s)      &   ($10^{40}$ erg/s)  &    timescale (s)  & (Mpc$^{-3}$ yr$^{-1}$) & (Mpc, \textbf{$z$})\\
\hline\\
SNIbc/II & 226     & $10^3$ & 1.0 & 500 & $10^{-3}$ & $2.8\times 10^3$, ~\textbf{0.50}\\
SNIa     & $2.1\times 10^{-5}$ & $10^2$ & 0.0 & 0.01 & $10^{-2}$ & 3.8,~~~~~~~~\textbf{0.0009}\\
LL-GRB   & 290     & $10^4$ & 0.0 & $10^4$ & $3\times 10^{-5}$ & $3.8\times 10^4$, ~~~\textbf{4.2}\\
TCE      & 0.0062  & $10^2$ & 0.0 & 10 & $10^{-4}$ & 120, ~~~~~~~~\textbf{0.028}\\
TDE      & 77.6    & $10^2$ & 0.0 & $5\times 10^5$ & $5\times 10^{-5}$ & $1.8\times 10^4$, ~~~\textbf{2.3}\\
SSS      & 1.1     & $10^{-4}$ & 0.0 & $5\times 10^5$ & 30 & 18, ~~~~~~~~~~\textbf{0.004}\\
ULX      & 411     & $1$ & 0.5 & $10^5$ & 0.1 & 920, ~~~~~~~~~~\textbf{0.19}\\
Type I bursts & 395 & $10^{-2}$ & 0.0 & 100 & 30 & 3.8,~~~~~~~~\textbf{0.0009}\\
 \hline
 \end{tabular}
% \end{center}
 \end{table*}

We simulated the WFXT performance for variable and transient sources, other than AGNs, assuming a simplified burst model where a source of luminosity $L_X^{quiesc}$ in the pre-burst phase, undergoes a burst of constant luminosity $L_X^{burst}$ for a duration equal to the characteristic timescale of the actual X-ray lightcurve. 
We computed the number of expected detections, requiring that a source is detected with $S/N>5$ and that its variable nature is ascertained with significance $S/N>3$ with respect to the pre-burst phase, i.e. the burst starts during the observation.
This simplified scheme allowed us to derive detection rates without the need of a specific knowledge of the characteristic of each source; obviously a more sophisticated approach is desirable and shall be implemented for each source separately in the future. The number of objects is calculated integrating over the cosmological volume accessible by WFXT for each source, given the parameters and constraints discussed above, and assuming an average volume density for the whole population. We did not explicitly include any evolutionary term, which is appropriate for most sources that will be observable only in the local Universe. 

Figure \ref{transients_histo} shows the number of detections in all surveys for the different classes discussed above: in particular hundred of LL-GRBs, XRF, TDE are expected over cosmological distances, mainly from the Deep survey. Recurrent bursters (ULX, Type I bursts) on the other hand will be mainly observable in the nearby Universe. For comparison we also show the predictions for the EXIST mission within 200 Mpc, as reported in \cite{Soderberg09}, even though it must be kept in mind that this mission will be using different energy bands and sampling patterns. 
In Table \ref{transient_tab} we report the input parameters and the numbers derived from our simulations. It must be stressed however that the properties of most sources considered here may span a wide range and/or be affected by large uncertainties. The values presented in this work are mainly intended to highlight the mission capabilities as a function of the observational parameters. An online version of the \textit{WFXT transient simulator} is available at \url{http://wfxt.na.infn.it/}, allowing the interested user to test additional type of source and/or parameter combinations.

\begin{figure*}[t!]
 \resizebox{1.07\hsize}{!}{\includegraphics{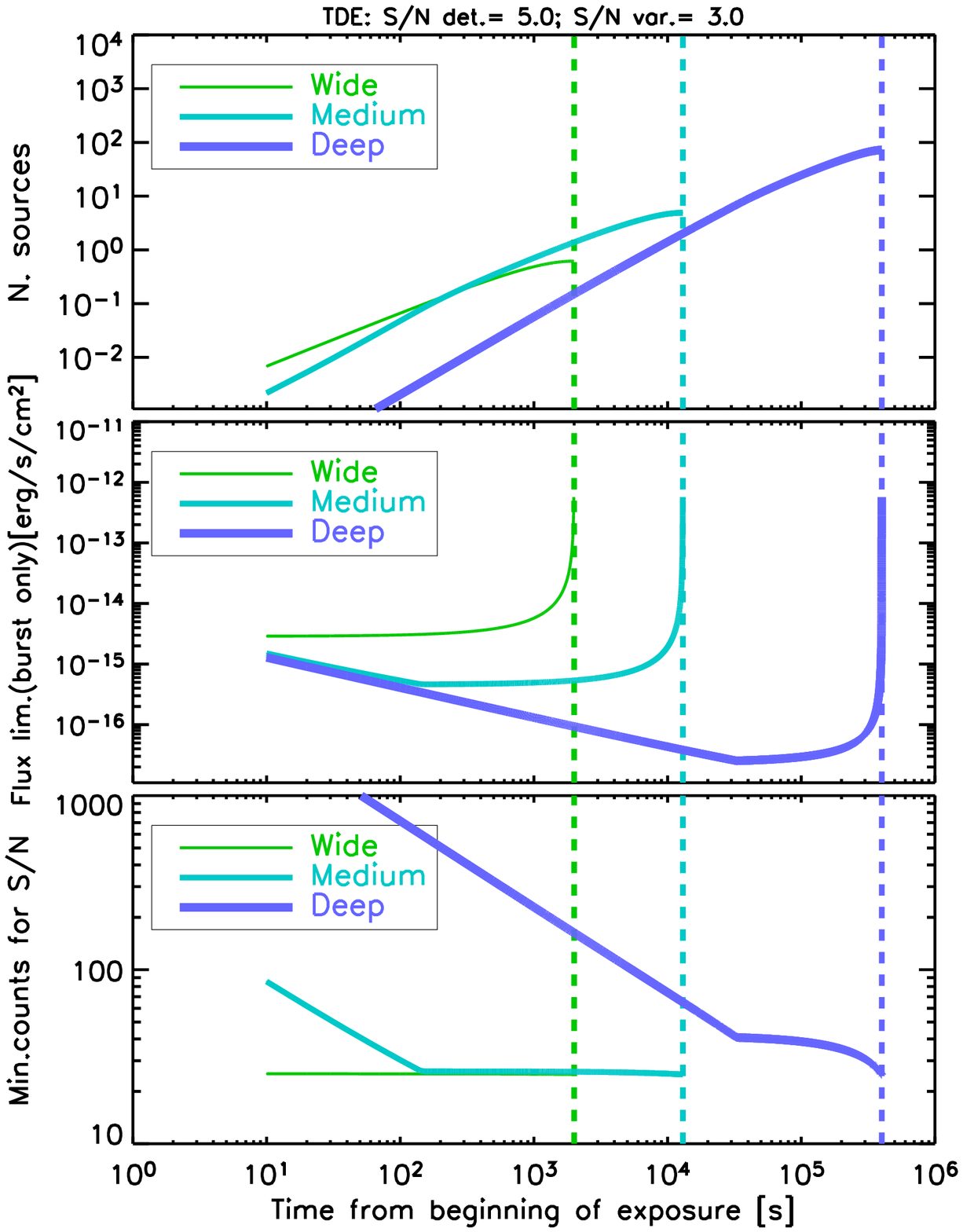}\includegraphics{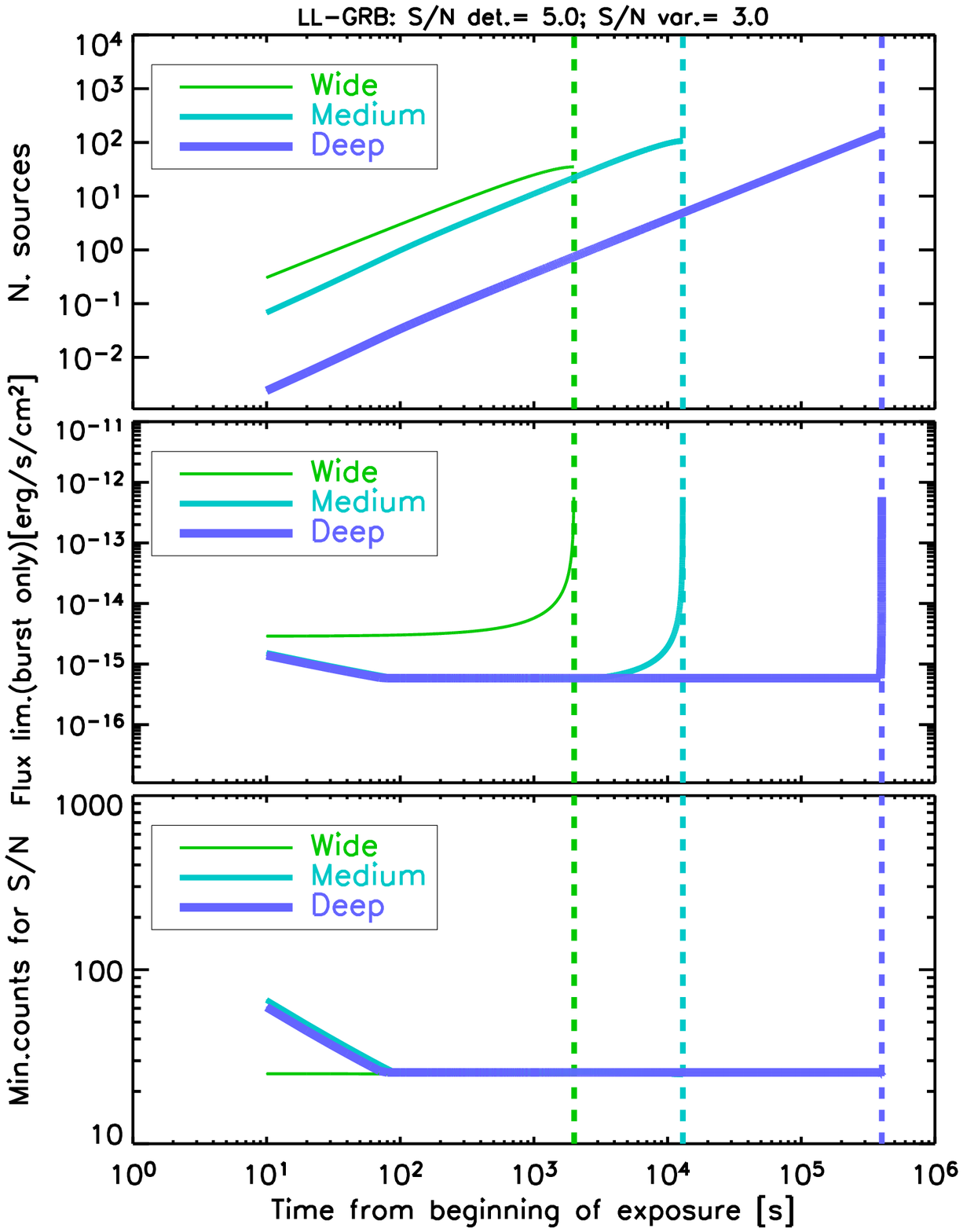}}
 \caption{
 \footnotesize Detection number and limiting fluxes and counts (tob, middle and bottom) for TDE (left panels) and LL-GRB (right panels) as a function of the sampled timescale. \label{transients}}
 \end{figure*}

The number of expected transients in each survey varies greatly depending both on the involved luminosities, volumetric rates and timescales. For instance frequent bursters (e.g. type I bursts) are likely to be observed mainly in the Wide survey, due to the greater area covered and despite the shorter exposure times. The interplay between these parameters can be observed in Figure \ref{transients}, where we show the number of detections, flux limit and minimum counts for TDEs and LLGRBs. The different behavior of the two classes is caused by the different luminosities and timescales, since the volumetric rates are comparable. In particular the duration of the burst, compared with the survey duration, affects the shape of the curves because it reflects our ability to verify the source variability, especially close to the flux limit. Also note that the WFXT low background allows to detect transients down to $\sim 25$ counts, but transients appearing early during the survey will end up having several thousand counts, thus allowing a good characterization of the lightcurve and of the spectral properties of the event. 

\section{Conclusions}
The time domain is rapidly opening up for astronomical studies, at all wavelengths. For instance in the near future a number of optical facilities, such as Pann-STARR and LSST, will allow to monitor the whole sky with unprecedented speed. While not a monitoring missions WFXT, with its large effective area, wide FOV and stable PSF, promises to offer a complementary view of the variable high-energy Universe. WFXT will allow to study thousand of variable AGNs, and hundred of other transient and variable sources. The study of such populations has been mostly limited to the local Universe so far, while WFXT will be able to sample cosmologically relevant volumes, thus constraining their rates, evolution and the underlying physical processes. 
Finally, recent studies suggest that the X-ray band could be the optimal energy range to use in order to identify triggers and/or counterparts for next generation Gravitational Wave and Neutrino experiments \citep[see, e.g.][]{Guetta10}; WFXT thus would prove extremely valuable in validating and characterizing the astronomical events detected by these facilities.
\begin{acknowledgements}
We wish to thank S. Gezari, L. Stella, P. Giommi, 
A. Cavaliere and A. Paggi for helpful discussions and comments.
MP and DdM acknowledge ASI financial support through contract I/088/06/0.
\end{acknowledgements}

\bibliographystyle{aa}

\end{document}